\documentclass[preprint,12pt]{elsarticle}
\usepackage{xcolor}
\usepackage{amssymb}
\usepackage{epsfig}
\usepackage{epstopdf}
\usepackage{graphicx}
\usepackage{amsmath}
\usepackage{newtxtext,newtxmath,wasysym}

\headheight=-.cm

\colorlet{darkgreen}{green!50!black}
\colorlet{brightyellow}{yellow!75!red}
\colorlet{orange}{red!50!yellow}
\colorlet{darkred}{red!80!black}
\colorlet{darkblue}{blue!50!black}

\usepackage{soul}

\journal{Physics Letters B}

\begin{document}

\begin{frontmatter}
\title{Antiproton-deuteron hydrogenic  states in optical models}
\author[IPHC]{R.~Lazauskas}
\address[IPHC]{IPHC, CNRS/IN2P3, Universit\'e de Strasbourg, 67037 Strasbourg, France}
\author[IJCLab]{J.~Carbonell}
\address[IJCLab]{Universit\'e Paris-Saclay, CNRS/IN2P3, IJCLab, 91405 Orsay, France}

\date{\today}

\begin{abstract}
By solving the Faddeev equations for the \={p}pn system, we
compute the antiproton-deuteron level shifts and widths  for the
lowest hydrogenic  states as well as the corresponding
\={p}d scattering lengths and volumes. The  \={p}d annihilation
densities are obtained and compared to the nuclear density of 
deuterium. The validity of the Trueman relation  for composite
particles is studied. The strong part of \={N}N interaction is
described by two different optical models,  including the
\={p}p-\={n}n coupling and n-p mass difference, while for NN
several  realistic interactions are used.
\end{abstract}

\vspace{-0.5cm}

\end{frontmatter}



\section{Introduction}

Low energy antiproton physics was an active field of research  at
the LEAR facility (CERN), from the beginning of the 80's up to its
closure in 1996. Among the rich variety of the experimental
program,  the formation and study of antiprotonic atoms in light
nuclei took an important place
\cite{BATTY_RPP52_1989,A_NPA658_99,REV_LEAR_KBMR_PREP368_2002,Gotta_PPNP52_2004,REV_LEAR_KBR_PREP413_2005}.
Antiprotons at rest  were captured  in highly excited atomic
orbits after ejecting one of the electrons. Then, these antiprotons
undertook a series of radiative and L-mixing Stark transitions in
consecutive cascades until reaching the lowest levels,  where
influenced by the strong \={N}N forces, and annihilating with the
nucleons. The detection of the X-ray from the the very last
transitions before annihilation (S- and P-waves) allowed to
extract the shifts and widths of the hydrogenic energy levels and
use them to parameterize the   \={N}N interaction \cite{REV_NNB_Frontiers_2020}.

The same Coulomb-like states are planned  to be used in the near
future in the framework of the PUMA project \cite{PUMA_LI_2017}
devoted to study the peripheral nucleons  in short lived unstable
isotopes. Indeed, the annihilation process may provide a unique
sensitivity to the neutron and proton densities at the
annihilation site, i.e. in the tail of the nuclear density, with
respect to more traditional probes.

Most of the theoretical analysis underlying the PUMA project are based on  antiproton-Nucleus (\={p}A) optical models since even
for the simplest \={p}p-\={n}n system (protonium),  calculations
are quite involved \cite{CIR_ZPA334_1989}. However we believe it could be of interest to obtain exact solutions for those light
nuclei which could be accessible by {\it ab-initio} techniques, that is with the only input of the NN and \={N}N interactions.

We present in this letter the first realistic solution of the
antiproton-deuteron (\={p}d) system, considered as a coupled
\={p}pn-\={n}nn three-body problem, based on realistic NN and
\={N}N interactions taking into account the deuteron D-wave, the
\={p}p-\={n}n charge exchange coupling and the p-n mass difference.


The \={N}N interaction is known to be strongly attractive
in most of partial waves  \cite{REV_NNB_Frontiers_2020}. As a consequence the three-body  $\bar
ppn$ system has a very rich spectrum of quasi-bound and resonant
states, which can be very close to \={p}-d threshold or
bound by  hundreds of MeV. Among them, there is an infinite
number of hydrogen-like  \={p}d states, just below
the deuteron threshold,  with energies
 \begin{equation}\label{epsilon_n}
  E_n\equiv E_d + \epsilon_n  \approx E_n^{(0)}\equiv E_d  + \epsilon_n^{(0)}  \quad\qquad \epsilon_n^{(0)}= -{R_y({\bar pd})\over n^2}
  \end{equation}
where $E_d$=-2.2246 MeV is the deuteron energy,  and $\epsilon_n^{(0)}$
are the energies  of an hydrogenic \={p}-d atom with a pointlike deuteron, and $R_y({\bar pd})$ its Rydberg
constant.  
These states correspond to an antiproton orbiting
around the deuteron in a Coulomb like orbit. They play a major
role in the context of  PUMA project and will be analyzed in
this work.

Our aim is to compute the energies and wave functions of these
states as well as the corresponding \={p}d scattering lengths.
From them we will obtain the \={p}d annihilation densities. Finally, we
 will investigate the
reliability of  the Trueman relation in the context of composite
systems. To solve the  \={p}pn 3-body problem we have used  the Faddeev equations (FE) in configuration space, which have produced 
in the past remarkable results in several branches of physics
\cite{LC_FBS31_2002,LC_PRC70_2005}. We present in this work a formal extension of the Faddeev formalism 
allowing to solve  coupled three-body systems, like \={p}pn-\={n}nn.

The main properties of the \={N}N optical models and the generalisation of the FE will be briefly described in the next Section \ref{Formalism},
the numerical results  will be presented in Section \ref{Results}
and some concluding remarks \ref{Conclusion} will close the paper.

\section{The formalism}\label{Formalism}



\subsection{The N\={N} interaction}\label{VNNB}

In the meson exchange picture of nuclear forces, the real part of the \={N}N interaction ($U_{\bar NN}$)  is obtained
as the G-parity transform of the NN one ($V_{NN}$).
Thus, if $V_{NN}$ is given by a coherent sum of different meson ($\mu=\pi,\rho,\omega,...$) contributions
one obtains $U_{\bar NN}$ by simply changing the sign of some of them according to
\begin{equation}\label{UNN}
 V_{NN}= \sum_{\mu}  V_{NN}^{(\mu)}   \qquad \Rightarrow \qquad
 U_{\bar NN}(r)  = \sum_{\mu} \;  {\rm G}(\mu)\; V_{NN}^{(\mu)}(r)
 \end{equation}
where G($\mu$) is the G-parity of meson $\mu$. G is related to the Charge-conjugation (C) and isospin (T) quantum number by  G= C$(-)^T$.
$U_{\bar NN}$ must be regularized below some cut-off radius $r_c$ in order to avoid the non integrable singularities
coming from the spin orbit and tensor terms.

The annihilation dynamics can be described either by explicitly
including annihilation  channels,  coupled to N\={N} in a unitary
way,  or by introducing an imaginary potential -- optical model
(OM) -- which accounts for the loss of the flux in the \={N}N
channel due to annihilation. We will adopt in what follows the OM
approach. In this case, the full  N\={N} interaction ($V_{N\bar
N}$)  is given as a sum of two terms
\begin{equation}\label{VOM}
 V_{N\bar N}(r)=  U_{\bar N N}(r)+  W(r)
 \end{equation}
where $W$ is a complex potential whose  parameters are adjusted
with experimental data. Several optical  models exist describing
well the bulk of low energy \={N}N physics
\cite{DR1_PRC21_1980,CLLMV_PRL_82,DR2_RS_PLB110_1982,NNB_CC_Nijm_1984, KW_NPA454_1985,PARIS_PRC79_2009,Haidenbauer_JHEP_2017}.
In this work we will use two of them : the meson-exchange inspired \={N}N  Kohno-Weise (KW)  potential   \cite{KW_NPA454_1985} 
and the  recently developed chiral EFT  J\"{u}lich model
\cite{Haidenbauer_JHEP_2017}. The last model keeps  only the
(G-parity transformed) pion-exchange contributions, whereas higher
momenta exchanges are represented by regularized contact
interaction terms with coupling constants adjusted to fit the
experimental data. These two models are quite different in
philosophy, but describe equally well the low lying protonium
states \cite{CIR_ZPA334_1989,Haidenbauer_JHEP_2017} and will
provide an estimation of the theoretical uncertainties in the \={p}d description.

\bigskip
If the isospin basis is commonly used to obtain the NN and  \={N}N
potentials, this basis is not well adapted to describe the low
energy  \={p}A physics where the Coulomb interaction play a
relevant role. Instead, we have used the so called
"particle basis" where the \={p}p and \={n}n states are coupled by
the charge exchange interaction and obeys the Schr\"{o}dinger
equation
\[ (E- H_0 ) \Psi= \hat{V}\;  \Psi    \quad \qquad \Psi=  \begin{pmatrix}   \Psi_{ p\bar{p}}   \cr \Psi_{n\bar{n}}  \end{pmatrix}    \quad
\hat{V}= \begin{pmatrix}
V_{p\bar{p}}     & V_{ce}      \cr
V_{ce}             &       V_{n\bar{n}}                 \end{pmatrix}
\]
$H_0$ a channel-diagonal kinetic energy term.
Using the isospin conventions \cite{Gasiorowicz_1966,CIR_ZPA334_1989}
\begin{equation}\label{Nbar_doublet}
N = {p\choose n}  \quad   \bar{N} = {- \bar n\choose + \bar p}   \equiv
\begin{array}{lcl}
|1/2,+1/2> &=& -|\bar n> \cr
|1/2,-1/2>  &=& +|\bar p>
\end{array}
\end{equation}
the particle basis is expressed in terms  N\={N}  isospin  states  $\mid T,T_3>$  as
\begin{equation}\label{particle_basis}
\begin{array}{lcl c lcl}
|p\bar{p}>&=&+{1\over\sqrt2}\left\{|00> + |10>\right\}  \\
|n\bar{n}>&=&+{1\over\sqrt2}\left\{|00> - |10>\right\}
\end{array}
\hspace{1.5cm}
\begin{array}{lcl c lcl}
|p\bar{n}>&=&-|1,+1>                                 \\
|\bar{p}n>&=&+|1,-1>
\end{array}
\end{equation}
and the  matrix elements of  $\hat{V}$  reads
\begin{equation} \label{Vpb_OM}
V_{p\bar{p}}=  {    V^0_{N\bar{N}} + V^1_{N\bar{N}}  \over2}   +
V_C   \qquad V_{n\bar{n}} = { V^{0}_{N\bar{N}} + V^{1}_{N\bar{N}}
\over2}+  2\Delta m \qquad
 V_{ce}        =  { V^{0}_{N\bar{N}} - V^{1}_{N\bar{N}}    \over 2}
\end{equation}
where  $V^T_{N\bar{N}}$  denotes to T  component of the \={N}N potential. $V_{n\bar{n}}$ incorporates the p-n mass difference $\Delta$m=$m_n$-$m_p$=1.293 MeV
 and $V_{p\bar{p}}$  the Coulomb  interaction $V_C$.
While the \={p}n states have isospin fixed to T=1,  the \={p}p
states involve isospin mixture and  always are coupled to \={n}n
ones. This coupling, as well as the $\Delta m$ term,  can have
sizeable effect on some nearthreshold \={p}p states
\cite{CIR_ZPA334_1989}.

\subsection{Faddeev equations}\label{FADEQ}

\bigskip
The $\bar{p}d$ system  is considered as a three particle system
$(\bar{p},p,n)$ interacting via pairwise potentials. We aim to
solve the quantum mechanical problem  using the Faddeev equation
in configuration space. However the  standard formalism
~\cite{LC_PRC70_2005} should be generalized to include the
\={p}p-\={n}n coupling.

The three traditional Faddeev  components (FC), corresponding to
$(\bar{p},p,n)$, must be supplemented with  three additional ones
corresponding to $(\bar{n},n,n)$.
\begin{figure}[h!]
\begin{center}
\epsfxsize=7.cm\centerline{\epsfbox{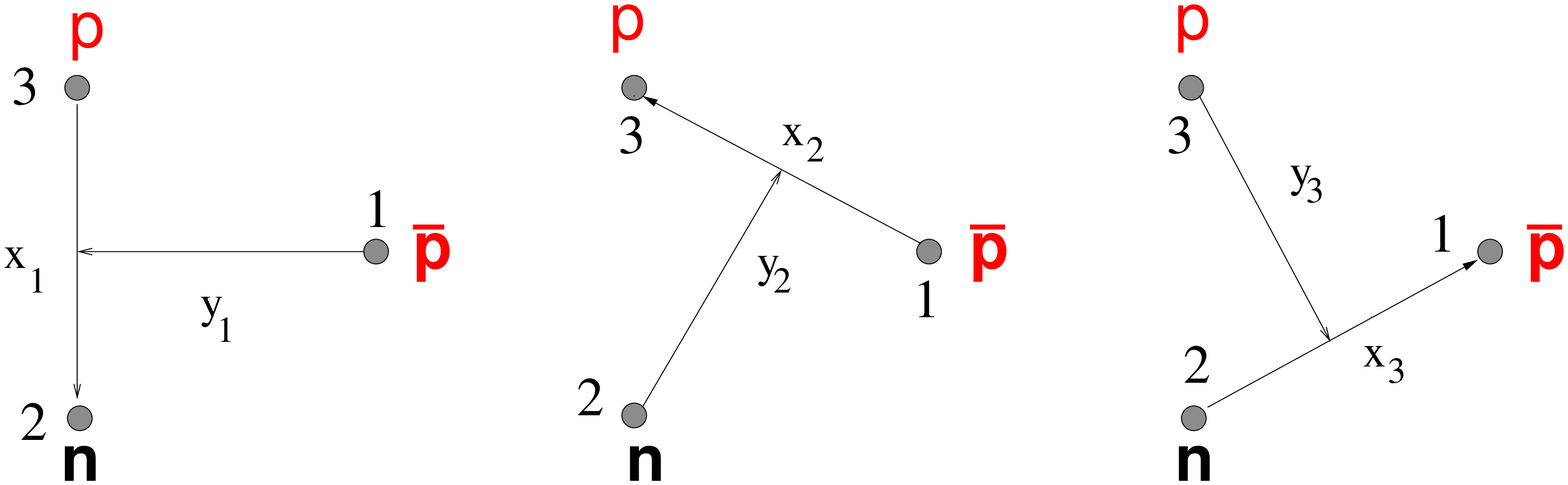}}
\vspace{0.5cm}
\epsfxsize=7.cm\centerline{\epsfbox{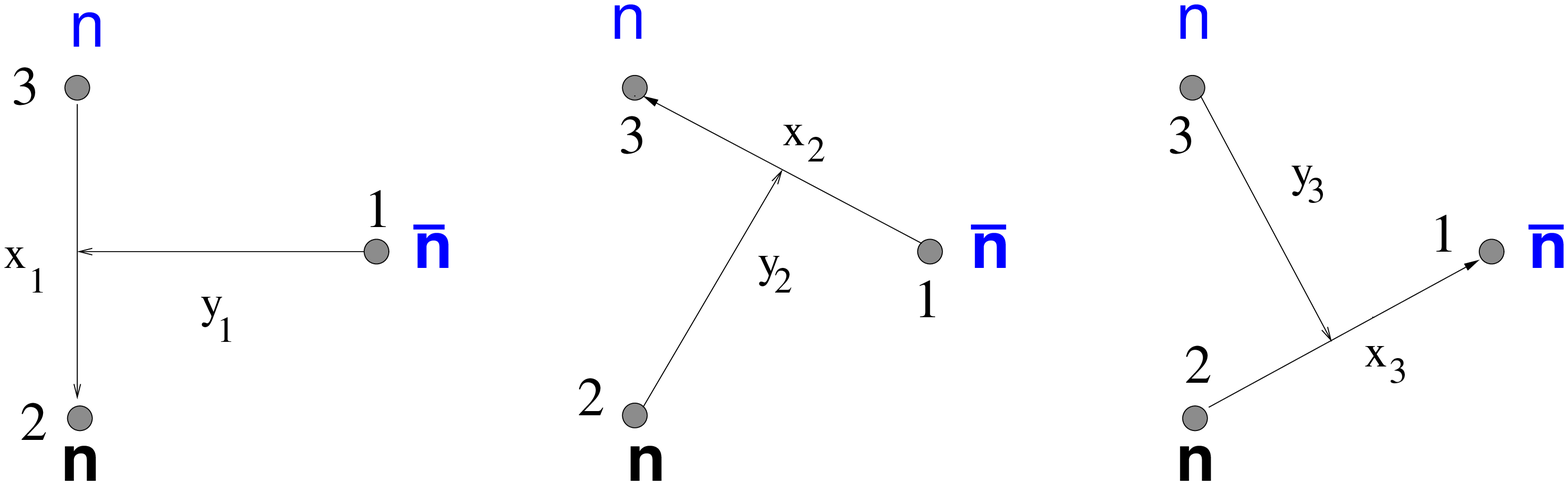}}
\end{center}
\vspace{-0.8cm} \caption{Faddeev components and Jacobi coordinates
for the solution of the \={p}d problem}\label{Jacobi}
\end{figure}

They  are represented in Figure \ref{Jacobi}, together with their
natural sets of Jacobi coordinates. The three traditional FCs are
depicted in the upper part of this figure and the three new ones in
the lower part. They can be cast in "charge-exchange" channel
doublets
\[           \hat\Psi_{Nn}\equiv {\Psi_{pn,\bar{p}}\choose\Psi_{nn,\bar{n}}} (\vec{x}_1,\vec{y}_1)   \qquad
             \hat\Psi_{\bar{N}N} \equiv {\Psi_{\bar{p}p,n}  \choose  \Psi_{\bar{n}n,n}}  (\vec{x}_2,\vec{y}_2)
\qquad  \hat\Psi_{n\bar{N}}\equiv  {\Psi_{n\bar{p},p} \choose
\Psi_{n\bar{n},n}} (\vec{x}_3,\vec{y}_3)      \] according to the
same configuration than in Fig. \ref{Jacobi}. Notice that in our
 case, with two identical particles in the $(\bar{n},n,n)$
subsystem, the lower components of $\hat\Psi_{\bar{N}N} $ and
$\hat\Psi_{n\bar{N}}$ are formally identical, since the total
wave function should be antisymmetric in the exchange of two $n$'s. They are  related by
\begin{equation}\label{Phi3_P_Phi2}
 \Psi_{n\bar{n},n}  (\vec{x}_3,\vec{y}_3)  = P^+ \Psi_{\bar{n}n,n}
 (\vec{x}_2,\vec{y}_2),
 \end{equation}
where $P^+$ is a standard  permutation operator (see \cite{LC_PRC70_2005}).

The total   wave function  of the system  $\hat\Psi$, obeys
the Schr\"{o}dinger equation
\begin{equation}\label{Schro}
 (E-H_0) \hat{\Psi}=\hat{V}  \hat{\Psi}    \qquad \hat{V}= \hat{V}_{Nn} +  \hat{V}_{\bar{N}N}+ \hat{V}_{n\bar{N}}
 \end{equation}
with the potential matrices
\[  \hat{V}_{Nn} = \begin{pmatrix} V_{pn}&0\cr 0&V_{nn}  \end{pmatrix}   \qquad   \hat{V}_{\bar{N}N}= \begin{pmatrix} V_{\bar{p}p} & V_{ce}\cr  V_{ce} & V_{\bar{n}n}  \end{pmatrix}
\qquad \hat{V}_{n\bar{N}} =\begin{pmatrix} V_{n\bar{p}}&0\cr0&V_{
n\bar{n}}  \end{pmatrix} \] It can be obtained in terms of  the
Faddeev components
\begin{equation}\label{Partition}
 \hat\Psi (\vec{x},\vec{y})=  \hat\Psi_{Nn} (\vec{x}_1,\vec{y}_1) + \hat\Psi_{\bar{N}N}(\vec{x}_2,\vec{y}_2) + \hat\Psi_{n\bar{N}} (\vec{x}_3,\vec{y}_3)
 \end{equation}
By inserting  (\ref{Partition}) in (\ref{Schro}) we obtained  the
corresponding  Faddeev equations
\begin{eqnarray}
(E-H_0-\,\hat{V}_{Nn} \, )          \; \hat\Psi_{Nn} &=&
\hat{V}_{Nn}      \;  ( \hat\Psi_{\bar{N}N} + \hat\Psi_{n\bar{N}}
)  \cr (E-H_0-\hat{V}_{\bar{N}N} ) \;\hat\Psi_{\bar{N}N} &=&
\hat{V}_{\bar{N}N} \; ( \hat\Psi_{n\bar{N}} + \hat\Psi_{Nn} )  \cr
(E-H_0-\,\hat{V}_{n\bar{N}}\,) \;\hat\Psi_{n\bar{N}}      &=&
\hat{V}_{n\bar{N}} \; ( \hat\Psi_{Nn}       + \hat\Psi_{\bar{N}N}
)  \label{EF_Nbd}
\end{eqnarray}

In terms of the channel components, and by using $(\ref{Phi3_P_Phi2})$, eq. (\ref{EF_Nbd}) results  into the following  system of coupled equations
\begin{eqnarray}
(E-H_0-V_{pn}  ) \Psi_{pn,\bar{p}}   &=& V_{pn}\;  ( \Psi_{\bar{p}p,n} + \Psi_{n\bar{p},p}         )     \label{EQp1}\\
(E-H_0  -  V_{nn}) \Psi_{nn,\bar{n}}          &=& V_{nn} \; (1+P^+)\Psi_{\bar{n}n,n}           \label{EQp2} \\
(E-H_0-V_{\bar{p}p}  ) \Psi_{\bar{p}p,n}   &=& V_{\bar{p}p} \; (\Psi_{n\bar{p},p} + \Psi_{pn,\bar{p}} )  + V_{ce} \;  [ (1+P^+) \Psi_{\bar{n}n,n} + \Psi_{nn,\bar{n}}  ]     \label{EQp3}\\
(E-H_0-V_{\bar{n}n}  ) \Psi_{\bar{n}n,n}   &=& V_{\bar{n}n} \;   ( P^+\Psi_{\bar{n}n,n} +  \Psi_{nn,\bar{n}} )  + V_{ce}\;  ( \Psi_{\bar{p}n,p} + \Psi_{pn,\bar{p}}  + \Psi_{\bar{p}p,n}  )       \label{EQp4}\\
(E-H_0 - V_{n\bar{p}}  ) \Psi_{n\bar{p},p} &=& V_{n\bar{p}}  \;
(\Psi_{pn,\bar{p}} + \Psi_{\bar{p}p,n}         )     \label{EQp5}
\end{eqnarray}
The sixth equation  turns to be identical than the fourth one  by permuting two neutrons and becomes redundant.

\bigskip
In order to solve eq. (\ref{EQp1}-\ref{EQp5}) for a given
$J^{\pi}$, each FC is expanded in partial waves
\begin{equation}\label{PWE}
\Psi_i (\vec{x}_i,\vec{y}_i) = \sum_{\alpha_i}    { \phi_{i,\alpha_i} (x_i,y_i) \over x_i y_i }    \mathcal{Y}_{\alpha_i} (\hat{x}_i, \hat{y}_i)
\end{equation}
where $\mathcal{Y}_{\alpha_i} $ are the generalized bipolar
harmonics including spin and angular momentum couplings; in the
last expression $\alpha_i={i,l_x,l_y,L,S,J}$ denotes all the
intermediate quantum numbers. The corresponding reduced radial
equations take the form
\begin{equation} \label{EFPW}
\sum_{\alpha'}  \hat{D}_{\alpha\alpha'} \varphi_{\alpha'}(x_{\alpha},y_{\alpha})=
\sum_{\alpha'\alpha''} v_{\alpha\alpha'}  (x_{\alpha})
  \int_{-1}^{+1} du_{\alpha} h_{\alpha'\alpha"}( x_{\alpha},y_{\alpha},u_{\alpha}) \varphi_{\alpha"} 
(x_{\alpha"},y_{\alpha"}),   \end{equation}
where  indices $(i,\alpha_i)$ are  regrouped into single one $\alpha$, $D $ is the differential operator
\[ \hat{D}_{\alpha\alpha'}=  \frac{\hbar^2}{m_\alpha} \left[ q^2+\partial^2_{x_{\alpha}} + \partial^2_{y_{\alpha}}  +{ l_{x_{\alpha}}(l_{x_{\alpha}}+1)\over x_{\alpha}^2}  +{ l_{y_{\alpha}}(l_{y_{\alpha}}+1)\over y_{\alpha}^2}  \right]\delta_{\alpha\alpha'}    -v_{\alpha\alpha'}(x_{\alpha}) \]
and $h_{\alpha\beta}$ are some integral kernels.
 The numerical solution is searched by expanding two-dimensional radial functions $ \phi_{i,\alpha_i}(x_{\alpha},y_{\alpha})$
on the bases defined on Lagrange-meshes~\cite{Baye_PR565_2015}. To
describe the functional dependence in variable $x_{\alpha}$ the
basis functions constructed from Lagrange-Laguerre quadrature
turns to be the most appropriate. The extension of the wave
function in this direction is determined by the size of the
deuteron.

The dependence on variable $y_{\alpha}$ is more tricky -- on one
side one should ensure internal variation of the wave function,
when antiproton penetrates the deuteron and all three-nucleons are
actively coupled. On the other hand the physical extension of the
wave function in $y_{\alpha}$ direction is determined by a size of
$\overline{p}-d$ Rydberg state, which largely exceeds the size of
the deuteron. Therefore special Lagrange quadrature meshes were
constructed to highlight this behavior of the systems wave
function on  variable $y_{\alpha}$. In particular, we were
splitting this quadrature in two separate domains, interconnected
by a boundary condition imposed by means of Bloch
operators~\cite{Baye_PR565_2015}.
The PW expansion (\ref{PWE}) included all the amplitudes with angular momentum $l_x\leq 5, l_y\leq 5$.

In our numerical calculations  we have used the values $\alpha$=1/137.0360,  $\hbar c$=197.327  MeV \;fm. 
The nucleon masses  appearing in the kinetic energy operator were taken  equal to $\frac{\hbar^2}{m_N}=41.4711$ MeV.fm$^2$,
which correspond to  the average physical values  m$_p$= 938.2721 MeV and m$_n$=939.5654 MeV.
With this parameters, the  Rydberg constant in (\ref{epsilon_n}) is
$R_y({\bar pd})=\frac{m_N}{3}\alpha^2=16.6662$ keV.

\section{Results}\label{Results}

The \={p}d (\={p}pn-\={n}nn) states are labelled by their total angular momentum and parity $J^{\Pi}$,
which  incorporates the intrinsic negative parity of \={p} and \={n} ($j^{\pi}=1/2^-$). 
However, since the hydrogenic states on which we are interested are close to the Coulomb states,
they are alternatively, though abusively, labeled by means of the corresponding spectroscopic notation $^{2S+1}$L$_J$.
Indeed,   the tensor and spin orbit  terms of the strong interaction in game, break the total spin S and orbital angular momentum $L$.
Still, this L- and S- breaking  is  produced by the short range nuclear interaction, whereas the wave function extends far beyond
this region and  it is in fact configurated by the  L- and S-conserving Coulomb force. 
As a consequence,  each \={p}d hydrogenic  state has a strongly dominant  L-component, which  justifies the use of the spectroscopic notation.

 
For S-waves there are two negative parity states: the doublet $J^{\Pi}$=1/2$^-$ and the quartet $J^{\Pi}$=3/2$^-$, denoted in spectroscopic  notation
$^2$S$_{1/2}$  and $^4$S$_{3/2}$, where the last state  contains a D-wave $^2$D$_{1/2}$ admixture. For P-waves we have  five
positive parity states $J^{\pi}$=1/2$^+_1$, 3/2$^+_1$, 1/2$^+_2$,
3/2$^+_2$,5/2$^+$, or  $^2$P$_{1/2}$ $^2$P$_{3/2}$  $^4$P$_{1/2}$  $^4$P$_{3/2}$ $^4$P$_{5/2}$ respectively.

\subsection{\={p}d level shifts and widths}\label{Sec_DE}

\bigskip
The energies $E_n$ of the bound  \={p}d states have a
negative imaginary  part are due to the  \={N}N annihilation.They  are written  as
\[ E_n= E_R+ i\,E_I =  E_R -i {\Gamma\over2}     \qquad E_I<0  \]
$\Gamma$ is the width due to the strong interactions and does not account for the "natural" electromagnetic width.
For hydrogenoic states, these energies $E_n$ are very close to the Coulomb pointlike values  $E_n^{(0)}$ defined in (\ref{epsilon_n}).
It is convenient to present $E_n$ in terms of the differences
\begin{equation} \label{Delta_n}
 \Delta E_n= E_n - E_n^{(0)}=  \epsilon_n- \epsilon^{(0)}_n  - i {\Gamma\over 2} \equiv \Delta E_R -   i {\Gamma\over 2}
 \end{equation}
This convention has the advantage to get rid of deuteron energies and be more sensitive to small variations. 
The real part of the $\Delta E$  ($\Delta E_R$)  is traditionally named "level shift", whereas level  width  ($\Gamma$) represents   half of negative imaginary energy part.

\begin{table}[h!]
\begin{center}
\begin{footnotesize}
\begin{tabular}{| l    |  r r r r      |  c  |  }\hline
                                &  MT13            &  AV18              &    INOY            &  I-N3LO      &     -$\epsilon_n^{(0)}$ (keV) \\\hline
S-waves                      & \multicolumn{4}{c   | }{ $\Delta E$ (keV)  }&   \\  
             $^2S_{1/2}$, n=1  & 2.251-1.0045i   &  2.147-1.0440i     & 2.214-0.99433i    &  2.209-1.0509i   &   16.6662     \\
             $^2S_{1/2}$, n=2  & 0.294-0.1406i   &  0.279-0.1454i     & 0.289-0.13892i    &  0.288-0.1468i    &  4.16655  \\
             $^2S_{1/2}$, n=3  & 0.088-0.0433i   &  0.084-0.0446i     & 0.087-0.04271i    &  0.086-0.0451i     &  1.85180      \\            \hline
P-waves                      & \multicolumn{4}{c   | }{ $\Delta E$ (meV)  }   & \\  
             $^2P_{1/2}$, n=2  & 49.1-258.0i&  -55.3-239.2i & -56.2-241.1i &  -58.5-244.0i       &    4.16655  \\
             $^4P_{1/2}$, n=2  & 24.4-194.8i&  200.2-186.4i & 200.2-188.2i &  200.3-186.1i     &   4.16655   \\
             $^2P_{1/2}$, n=3 &  16.1-90.6i &  -14.0-83.94i  & -14.2-84.57i &  -15.0-85.61i       &   1.85180  \\
             $^4P_{1/2}$, n=3 &  8.62-68.4i &  59.4-65.51i  & 59.0-66.14i  &  58.4-65.36i          &   1.85180   \\ \hline
\end{tabular}
\end{footnotesize}
\caption{Complex \={p}d energy shifts $\Delta E_n$ obtained for
different NN interactions and the KW  \={N}N model. 
}\label{Tab_Delta_KW}
\end{center}
\end{table}

The  $\Delta E_n$ values  for the lower S- and P-waves  \={p}d states   are given in Table  \ref{Tab_Delta_KW},
together with the pointlike deuteron Coulomb levels $\epsilon_n^{(0)}$.
The have been obtained with  the same \={N}N model (KW) and 
different NN potential:  the phenomenological S-wave MT13  \cite{MT_NPA127_1969} and the realistic potentials AV18  \cite{AV18_1995}, INOY 
 \cite{Doleschall_PRC69_2004} and N3LO \cite{N3LO_EM_PRC68_2003}.
 The  simplistic MT13 interaction was chosen to estimate the role of deuteron D-wave and corresponding quadrupole moment in the final results.

\bigskip
For  L=0  states, the effect of strong \={N}N force is a $\approx 2$ keV shift in the ground state,   what represents 15\%
of its value. It is a quite remarkable effect in view of the large extension of these states ($R\equiv\sqrt{<r^2>}$= 75, 280,
620 fm respectively) compared to the range of strong  interaction ($\sim$ 0.7 fm)  and an indication of the \={N}N force strength.

Although $V_{\bar N N}$  is  strongly attractive in S-wave, the
global effect of  \={N}N force is repulsive (diminish the binding
energy or equivalently has $\Delta E>$0). This  effective
repulsion is produced by the strong suppression of the \={p}d
wave function near the origin,  due to the imaginary part of the
optical potential which "pulls out"  the energy levels towards the
continuum. The widths of these states $\Gamma$ = -2 Im($E_I$) turn to be of the same order  than the shifts.

One can see a nice agreement between the four NN models that we have considered, both for the real as well for the imaginary parts
of the energy, with  differences  of the order of 1\%. 
This independence of a 3-body result with respect the NN interaction  is  quite remarkable, in view of  the large differences existing  among them in the 3N observables.
For instance in the 3N binding energies,  which probe the off-energy shell structure of NN models, they may differ by as much
as 10\%. On another hand the \={p}d level shifts and width are determined by the short range part
of the wave function, a region where the deuteron wave functions themselves  can sizeably differ \cite{Deuterons}.
It might be explained by the strong suppression of the short range
wave function, keeping the three particles apart and so minimizing the differences related  to the 3-body off-energy shell effects.

\bigskip
For L=1 states, the level shifts $\Delta E_R$  are only few tens of meV in the  ground states,
and are  one order of magnitude large for their respective widths. This represents a very small variation relative to pointlike Coulomb values.

The P-level shifts of the three realistic NN models are also in a very good  agreement,  but those obtained with the
phenomenological MT13 interaction are totally different, even in sign. 
One can see that the absence of NN tensor force induces a very small hyperfine structure:
the $^2$P$_{1/2}$ level shift, e.g., vary from 49 to 24 meV with MT13
while it changes from -55 to +200 meV with AV18, say 25 meV compared to 250 meV, i.e. one order of magnitude. 
Despite that the MT13 energy shifts of the individual levels  differ significantly from the realistic model predictions,
their spin-averaged values are in  nice agreement (see Table \ref{Tab_aver}).

\begin{table}[h!]
\begin{center}
\begin{footnotesize}
\begin{tabular}{|l |  r r| r r |  }\hline
                                             &  \multicolumn{2}{c|}{ I-N3LO   +KW }      &  \multicolumn{2}{c|}{ I-N3LO   +J\"{u}lich }          \\
                                           &  \={p}p      & \={p}p + \={n}n & \={p}p  & \={p}p + \={n}n   \\\hline
             $^2S_{1/2}$, n=1 (keV)&  2.179-1.024i  &  2.209-1.050 i    &   2.028-0.928i    & 2.108-1.085i            \\
             $^2S_{1/2}$, n=2  (eV) &  284-143i        &  288-147 i         &     264-128i      &   274- 151i          \\
             $^2S_{1/2}$, n=3 (eV) &   85.3-43.9i      &  86.4-45.1 i       &     79.1-39.3     &   82.0-46.3i         \\\hline
 
             $^4S_{3/2}$, n=1  (keV)& 2.206-0.970i  &  2.306-1.045i   & 2.027-0.916i      & 2.321-1.216i      \\
             $^4S_{3/2}$, n=2 (eV) &  288-136i         &  302-147i       & 264-127i          & 302- 171i        \\
             $^4S_{3/2}$, n=3  (eV) &  86.6-41.7i      &  90.7-45.2i     & 79.1-38.8         & 90.7-52.6i       \\\hline

             $^2P_{1/2}$, n=2 (meV) & -61.6-210i    &  -58.5-244 i     & -105-194i         & 18.7-329i            \\
             $^4P_{1/2}$, n=2 (meV) &  214-158i     &   200-186 i      & 200-124i          & 171-194i              \\
             $^2P_{1/2}$, n=3 (meV) & -16.3-73.8i   &  -15.0-85.6 i    & -31.9-68.3i       & 13.2-120i             \\
             $^4P_{1/2}$, n=3 (meV) &  63.5-55.5i   &   58.4-65.4 i    & 59.1-43.5i        & 47.0-63.7i             \\ \hline

             $^2P_{3/2}$, n=2  (meV) & -60.3-201i    &  -76.2-226i     & -81.2-144i    &   -108-207i        \\
             $^4P_{3/2}$, n=2  (meV) &  43.6-180i    &   35.0-191i     & 55.0-137i    &    40.4-160i          \\
             $^2P_{3/2}$, n=3  (meV) & -17.3-68.6i   &  -21.4-79.5i    & -23.3-50.6i &  -32.7-72.7i          \\
             $^4P_{3/2}$, n=3  (meV) &  13.8-63.2i   &   10.7-67.0i    & 17.8-48.3i &   12.7-56.3i          \\ \hline
            $^4P_{5/2}$, n=2  (meV) &  57.6-185i    &  34.7-208i      & 7.1-132i    &  -21.6-205i               \\
            $^4P_{5/2}$, n=3  (meV) &  18.7-64.8i   &  10.7-72.9i     & 1.1-46.2i  &   -9.1-72.1i         \\ \hline
\end{tabular}
\end{footnotesize}
\caption{Complex level shifts  (\ref{Delta_n}) of atomic \={p}d states calculated with the same I-N3LO NN interaction (for deuteron)  and two  different \={N}N models:  KW \cite{KW_NPA454_1985} and Julich \cite{Haidenbauer_JHEP_2017}.}\label{Tab_kw_jul}
\end{center}
\end{table}

In order to  have a first hint on the  theoretical uncertainties due to \={N}N interaction,  it is interesting to
compare the KW predictions from Table \ref{Tab_Delta_KW}  with those provided by the J\"{u}lich model 
\cite{Haidenbauer_JHEP_2017}~\footnote{The original parametrization of \={N}N by J\"{u}lich group was made employing
relativistic kinetic energy operator. Such an operator can not be used in our calculations, and therefore some coupling constants of
J\"{u}lich \={N}N interaction were slightly readjusted to retain the same \={p}p  scattering lengths with a  non-relativistic kinematics.}, 
while using for both the same $V_{NN}$  interaction (I-N3LO).
The results  are  presented in table \ref{Tab_kw_jul}. The  full results are in columns  \={p}p-\={n}n, whereas  columns denoted
by \={p}p contains the values obtained by neglecting  the \={p}p-\={n}n coupling in the \={N}N interaction.

\bigskip
For S-wave, the  agreement between these two models  (\={p}p-\={n}n column) is reasonably good, with differences at the level of 5\%. 
This stability is also remarkable taking into account the large uncertainties in the \={N}N interaction and the very different background of the two models considered.

The energy shifts for P-waves are strongly \={N}N model dependent.
Some level shifts  differ even in sign, and some widths by 50 \%. These differences,  that occur also  in protonium
\cite{CIR_ZPA334_1989}, may be due to the existence of nearthreshold
P-wave singularities in the \={N}N scattering amplitude. A small
variation in the potential -- or an approximation in the
calculation -- can move a loosely bound state into a resonance,
what produces a change of sign in the level shifts. For the KW
model, these singularities were  analysed in \cite{CDPS_NPA535_91}
but  in fact they are present in most of the \={N}N models
\cite{SH_PR35_1978,DR_AP121_1979,PARIS_PRC79_2009}. Without an ad
hoc adjustment of these singularities at the level of
$V_{\bar{N}N}$, any coincidence will be fortunate. On the other
hand the experimental knowledge of the P-waves protonium level shifts  is too poor to attempt solving this issue.

The coupling to the \={n}n channel plays  a moderate role in the KW  model: 5\% in S-waves and 20\% in P ones. However the {p}p+\={n}n coupling is much stronger in the J\"{u}lich  interaction and becomes crucial in determining the  P-wave energy shifts. 
For instance, in  he  $^2P_{1/2}$ (n=2) state, $\Delta E$ changes from -105-94i meV to +18-329i meV, and similar changes happen for  $^4P_{5/2}$. 
It is thus essential to take properly into account the  \={p}p+\={n}n coupling if we aim to provide  the real predictions of a \={N}N model.

It is worth noticing the  strong hyperfine structure manifested with both \={N}N models between the $^2$P$_{1/2}$ and $^4$P$_{1/2}$ states and, mainly, 
 the $^2$P$_{3/2}$ and $^4$P$_{3/2}$ ones. It is a direct consequence of the NN tensor force and, as we have already pointed out,  is  strongly suppressed  
 when using the S-wave MT13 potential.

Finally, one may  remark that the above presented level shifts fulfill rather well  the  $ \sim 1/(n+L)^3$ scaling law for the
excited states of the same symmetry, as suggested by the Trueman relation \cite{Trueman_NP26_1961,CRW_ZPA343_1992}.



\subsection{Comparison with previous calculations and with experimental results}

\bigskip
There has been  in the past several attempts to compute   the
\={p}d level shifts
\cite{WGN_PLB_85,LT_PRC42_1990,YKK_PLB659_2008}\footnote{To avoid possible confusions with
the literature, it is worth noticing that in some papers the level shifts  $\Delta E_R$ is denoted by $\epsilon$,
whereas in some other experimental   \cite{Gotta_NPA660_1999} and theoretical  \cite{YKK_PLB659_2008} works the opposite sign convention -$\epsilon$ is used.}. 
All of them contain approximations both in the solution of the three-body
problem as well as in the NN and \={N}N dynamics: for instance they were limited to deuteron S-wave  and neglected the \={p}p-\={n}n coupling. 
In their spirit, these calculations are close to our \={p}p MT-13+KW model. It is worth to review them and compare with our findings.

The pioneering results of \cite{WGN_PLB_85} were based on a multiple scattering expansion. They include the single \={p}p
channel,  S-wave deuteron, and  uses a separable version of DR potential (model 1)  \cite{DR1_PRC21_1980}.  For  protonium the
last model provides very similar results as KW potential \cite{DR2_RS_PLB110_1982,CIR_ZPA334_1989}. For the S-wave   their
best values are $\Delta E(^2S_{1/2})$=  -2.14 - 0.59i keV and $\Delta E(^4S_{3/2})$= -2.19 - 0.64i keV. Their level shifts
are comparable with ours (see Tables \ref{Tab_Delta_KW} and \ref{Tab_kw_jul}) but the widths are by a factor two smaller.
Their P-wave level shifts and half widths,  listed in Table \ref{Tab_Comparison_P},  are almost degenerate  as a result of  S-wave deuteron in use. 
On the contrary, the estimated widths remain within a 20\% from our values obtained using MT-13+KW model and neglecting \={p}p-\={n}n coupling.

\begin{table}[h!]
\begin{center}
\begin{tabular}{ |l |    r r r| } \hline
             & MT13+ KW      & Ref. \cite{WGN_PLB_85}     &  Ref. \cite{YKK_PLB659_2008}   \\  \hline
$^2P_{1/2}$  &    32.3-185i  & 69 -199i     &  -99 -328i    \\
$^4P_{1/2}$  &    48.9-204i  & 60 -256i     & -101 -393i   \\
$^2P_{3/2}$  &    32.0-186i  & 66 -193i     &  -98 -322i   \\
$^4P_{3/2}$  &    37.4-193i  & 42 -215i     &  -97 -324i  \\
$^4P_{5/2}$  &    49.1-192i  & 41 -210i     & -101 -330i  \\\hline
\end{tabular}
\end{center}
\vspace{-0.5cm} \caption{Comparison of the P-waves complex energy
shifts $\Delta E$. Our results with KW and MT13 \={N}N models are
compared with previous calculations for Wycech et al.
\cite{WGN_PLB_85} and Yan et al
\cite{YKK_PLB659_2008}}\label{Tab_Comparison_P}
\end{table}

A second group  \cite{LT_PRC42_1990}  obtained  the \={p}d S-wave level shifts
 from a projected form of the Faddeev equations with rank-one, separable, S-wave two-body potentials  and the \={N}N  parameters fitted to  Graz model.
 This work contains several results depending on the approximation used.
 Their $^2$S$_{1/2}$ level shifts range from 1,38 to 1.48  keV and their half widths from 0.45 to 0.64 keV.
For the $^4$S$_{3/2}$  level shifts are in the interval [1,43,1.72] and corresponding widths in [0.36,0.42] keV. No  any result for P-waves were given.

The only  comparison with the previous results using the same KW potential is with Ref. \cite{YKK_PLB659_2008}. Starting from the
3-body Schr\"{o}dinger equation, these authors built an effective $\bar{p}-d$ two-body  interaction  and solved the corresponding
equation by means of Sturmian functions. Their solution is  based on an undistorted deuteron core with a purely S-wave deuteron wave function. 
For the S-waves,  they obtained  $\Delta E(^2S_{1/2})$=2.478 - 1.225 i keV and $\Delta E(^4S_{3/2})$= 2.503 - 1.235 i keV.
By doing so, the real parts differ by 10\%  from our  $\Delta E(^2S_{1/2})$ values quoted in Table \ref{Tab_kw_jul}, while the
imaginary parts are  by 20\% larger. The P-waves, given in Table \ref{Tab_Comparison_P}, differ significantly from our MT13+KW
results both in the level shifts -- which have even opposite signs - as well as in their respective widths.

\begin{figure}[h!]
\begin{center}
\centering\includegraphics[width=7.cm]{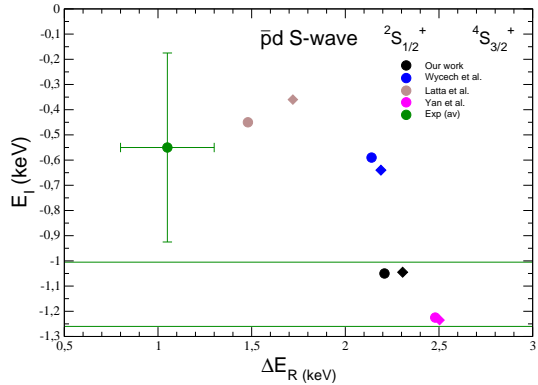}
\end{center}
\vspace{-0.5cm}
\caption{Comparison for  the \={p}d S-wave level shifts and width with previous results \cite{WGN_PLB_85,LT_PRC42_1990,YKK_PLB659_2008}  are plotted. } \label{Comparison_S}
\end{figure}

The ensemble of these results for  S-wave  is summarised in Figure \ref{Comparison_S}.
Filled circles correspond to $^2$S$_{1/2}$ and diamond to $^4$S$_{3/2}$. Among our results we have chosen those with N3LO+KW
models from Table  \ref{Tab_Delta_KW} and represented in black symbols,
blue symbols  correspond to Ref. \cite{WGN_PLB_85},   brown symbols to the results labelled 'Faddeev model' from  Table I of Ref \cite{LT_PRC42_1990}
 and  results from Ref. \cite{YKK_PLB659_2008} -- having changed the sign of the real part --  are in magenta.
The experimental values (green symbols) are taken from \cite{Gotta_AIP793_2005,Augsberger_PLB461_1999} and are S-averaged.
The horizontal lines correspond to the estimated width from  \cite{Gotta_NPA660_1999} where no level shifts were given.

 \vspace{0.5cm}

The results displayed in Figure \ref{Comparison_S} deserve some comments.  If one could explain the dispersion among the
theoretical results in terms of approximate solutions, our results
are exact in the numerical sense and the difference among the KW and J\"{u}lich models is not significant. Nevertheless  our
predictions are in strong conflict with the spin averaged experimental values, in particular for the level shifts which
differ by  a factor 2. These are spin average S-wave values but
notice that $^2$S$_{1/2}$ (filled circles) and  $^4$S$_{3/2}$
(filled diamonds) are almost degenerate in all the calculations.
Since such a discrepancy is not observed in protonium, it would be
interesting to have an experimental confirmation of the \={p}d S-wave level shifts to clarify this point. The S-wave level shifts
and widths probe the short range part of the interaction and is a privileged path to access the strong and annihilation \={N}N forces.

\bigskip
Concerning the  \={p}d experimental results one has not   been
able yet to disentangle the hyperfine splitting of atomic levels.
These levels are indeed strongly overlapping due to their large
widths, complicating their differentiation. However, the weighted
average values of level shifts and widths for S and P  states have
been measured at LEAR
\cite{Augsberger_PLB461_1999,Gotta_NPA660_1999,Gotta_AIP793_2005}.
The comparison  of our spin-averaged calculated values for L=0
and L=1 states is made in Table \ref{Tab_aver}. The theoretical
predictions for S-states are very stable with respect to all the
considered  NN and \=NN potentials. If the width falls inside the
experimental error it is indeed quite worrying that the level
shif represents almost double of the measured ones. The
theoretical discrepancies in the 2P level shifts is much larger,
and differ by one order of magnitude from the experimental values.
Nevertheless, in this case, the spread between the model predicted
values is also significant. One may hope that this discrepancy
with phenomenology might be resolved in the  near future, once
more accurate \=NN data will allow  to fix the  \=NN interaction
parameters in P-waves. Any experiment aiming to identify the
 bound or resonant states in \={N}N P-waves would be more than welcome.

\begin{small}
\begin{table}[h!]
\begin{center}
\begin{tabular}{|l |  c c c c c | c | c|} \hline
                                               &  MT13       &  AV18      &    INOY      &  I-N3LO & I-N3LO & Ref.  \cite{WGN_PLB_85} & Exp. \\
                                               &  +KW        &    +KW      &    +KW      &  +KW & +J\"{u}lich &  & \\\hline
L=0  $\Delta$E(eV)                &     2297     &   2194           &   2268              &     2274   &  2250  &  2170  & 1050$\pm$250  \cite{Augsberger_PLB461_1999,Gotta_NPA660_1999,Gotta_AIP793_2005}    \\
L=0  $\Gamma$ (eV)             &     1982     &    2129           &   1971              &     2095   &  2344  &  1250  & 1100$\pm$750  \cite{Augsberger_PLB461_1999,Gotta_NPA660_1999,Gotta_AIP793_2005}  \\
                                               &                  &                   &                     &            &        &        & 2270$\pm$260   \cite{Gotta_NPA660_1999}           \\
L=1  $\bigtriangleup$E (meV)&       26.6    &    22.5          &      20.7    &     18.2   &  -1.1  &  52    & 243$\pm$26   \cite{Gotta_NPA660_1999}   \\
L=1  $\Gamma$ (meV)          &      428      &   414 & 420
&     420    &  416   &  422   & 489$\pm$30
\cite{Gotta_NPA660_1999}  \\         \hline
\end{tabular}
\caption{Spin-averaged level shifts ($\Delta E_R$) and widths
($\Gamma$) compared to  LEAR experimental results}\label{Tab_aver}
\end{center}
\end{table}
\end{small}

\subsection{Annihilation densities}\label{AD}

The \={p}d  complex energies $E_n$ are obtained as eigenvalues of
the FE and control the asymptotic behavior of the solution. They
can alternatively be obtained in terms of the total \={p}pn
wavefunction in an integral form
\[ \Delta E = E_R - i \;{\Gamma\over 2} = < \Psi_{\bar pd} \mid H_0 + V  \mid \Psi_{\bar pd} >  \]
what constitutes a robust test for the calculations. In
particular, taking the non Hermitian part of this expression
provides an integral form for $\Gamma$,
 in terms of the imaginary part of the optical potential (\ref{VOM})
 \begin{equation}
{\Gamma}  = 2 \int_0^{\infty} y^2 dyx^2dx \; {\rm Im}[ W(x,y)] \mid  \Psi_{\bar{p}d} \mid^2    = \int_0^{\infty}   dr\; \gamma_a(r)
\end{equation}
where $r$ is de distance between  \={p} and  deuteron center of
mass ($r= {\sqrt{3} \over2}y$ with our conventions for the Jacobi
coordinates (see Fig. \ref{Jacobi}). The function  $\gamma_a(r)$,
named   annihilation density,  can be interpreted as the
probability for an \={p} to be annihilated with a proton or with a
neutron at a distance r from the center of mass of the target
nucleus.

\begin{figure}
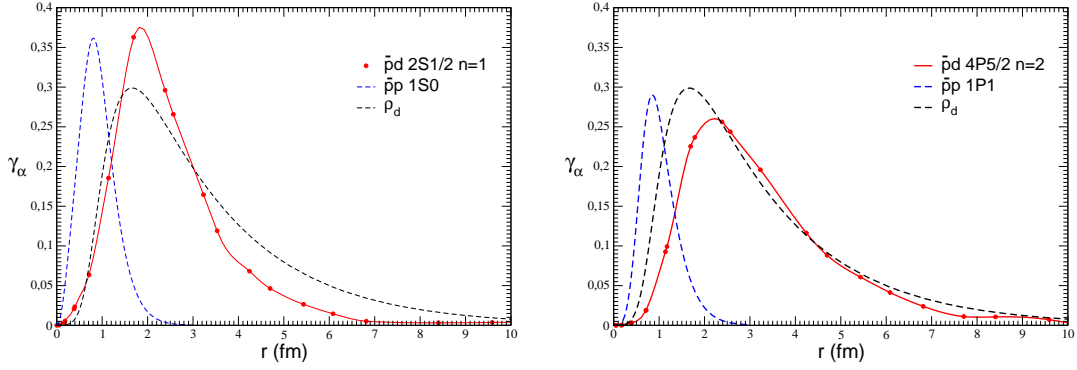

\begin{center}
  \includegraphics[width=0.4\textwidth]{pbd_AD_2S12_n_1.eps}\hspace{0.5cm}
  \includegraphics[width=0.4\textwidth]{pbd_AD_4P52_n_2.eps}
\caption{\={p}d annihilation densities $\gamma_a$ for the 2S1/2
(left panel) and 4P5/2 (right panel) states calculated with the
MT13+KW model.
They are compared with the  \={p}p $^1$S$_0$ and $^1$P$_1$ $\gamma_a$'s
in protonium  and with corresponding deuteron matter density $\rho_d$.}\label{Fig_ann_density}       
\end{center}
\end{figure}

One of the aims of our work was to obtain the \={p}d annihilation
density $\gamma^{\bar pd}_a$ and compare it with the deuteron
density $\rho_d$. It is also interesting to compare the same
quantity in absence of any matter density, that is with the
\={p}p (protonium) case ($\gamma^{\bar pp}_a$), considering p as
point-like particles \cite{CIR_ZPA334_1989}. This is done in
Figure \ref{Fig_ann_density} for  \={p}d  S- and  P-states.
 We have arbitrarily normalized the different quantities  -- $\gamma^{\bar pd}_a$, $\gamma^{\bar pp}_a$ and $\rho_d$ -- for the sake of comparison.

For the  $^2$S$_{1/2}$ (left panel),  $\gamma^{\bar pd}_a$ (solid
red line) is peaked around r=2 fm but the comparison with the
$^1$S$_0$ $\gamma^{\bar pp}_a$ (dashed blue line) tell us that this
process is driven by  the  deuteron density (dashed black line)
since a substantial part of it takes place at $r>2$. An even nicer
picture is obtained for the  $^4$P$_{5/2}$ state: the absorbtion
density maximum is situated at $r\approx3$ fm  and scales nicely
with the peripheral  (r$>$3 fm) deuteron density. On the contrary,
the protonium $^1$P$_1$ annihilation is centered at around $r=0.8$ fm.


One may conclude that in the P-states antiprotons are absorbed
mostly at the surface of the nucleus, thus confirming the intuitive
conjecture on which the PUMA project, aiming to study peripheral
neutron densities in the exotic nuclei, is based. On the contrary in
S-wave antiprotons penetrate easily into the nucleus and their
annihilation is less peripheral. Finally, for L$>$1 angular
momentum states annihilation should be even more peripheral.
Therefore the comparison of the annihilation products from different
angular momentum states could provide a hint on the  radial evolution of
the neutron/proton ratio in the target nucleus.


\subsection{\={p}d scattering lengths and Trueman relation}\label{Full}

By adding the corresponding inhomogeneous term in  the Faddeev equations (\ref{EQp1}-\ref{EQp5})  we have calculated the
 \=p-d scattering "lengths" $a_L$\footnote{Although traditionally  named "scattering volume" for L=1 we keep abusively the
 "scattering length"  denomination for all L states.}.
 Although these values could be approximately obtained, as in the protonium case \cite{CRW_ZPA343_1992},
 from the complex level shifts $\Delta E_n$ by using the Trueman relation \cite{Trueman_NP26_1961},
 we preferred to proceed in the reverse order. That is,  insert in Trueman
 relation the computed $a_L$ values and compare the  $\Delta E_n$  thus obtained
  with the exact $\Delta E_n$ values displayed in Tables  \ref{Tab_Delta_KW} and \ref{Tab_kw_jul}.

Indeed, Trueman  demonstrated in \cite{Trueman_NP26_1961}  that the low energy  parameters  of two opposite-charged particles
are related to the energy-shifts of their atomic states according to an expansion
 \begin{equation}\label{Trueman}
{\Delta E_n\over E^c_n}  \approx -{4\over n}\alpha_{nL}  {a_L\over B^{2L+1} }   \left[   1 -    \beta_{nL}  {a_L\over B^{2L+1} }  \right]   + ...
 \end{equation}
where $a_L$ are the Coulomb corrected scattering "lengths", B is
the corresponding Bohr radius, and $\alpha_{nl},\beta_{nL}$ some
numerical coefficients
 (see  \cite{CRW_ZPA343_1992} for details).
We have shown in \cite{CRW_ZPA343_1992}   that these relations
turn to be very accurate for \=pp atomic states. It is not obvious, however, how well these relations hold for a \={p}
scattering on a composite nucleus, where the nuclear charge is
distributed over a significant volume. Furthermore  if it is not
 spherically symmetric, thus giving rise to long-range
Coulomb multipole terms and breaking  the conservation of spin
quantum number. Indeed, due to the presence of quadrupole moment
in deuteron, the effective \=p-d interaction has an  asymptotic $1/r^3$ terms 
beyond the attractive Coulomb term.

The accuracy of  expansion (\ref{Trueman}) for the  \=p-d system is shown in Table
\ref{Tab_Trueman}  with  two combinations of NN and \={N}N models.
The computed scattering "lengths" $a_L$ are given in the first
column, the $\Delta E$ values obtained by expansion
(\ref{Trueman}) are given in  columns $\Delta E_1$ and $\Delta
E_2$, representing respectively corrections up to the first and
the second order. Column $\Delta E$ represent  the exact values
obtained via direct bound state calculation and imported from 
Table \ref{Tab_Delta_KW}.

These results demonstrate that Trueman expansion  (\ref{Trueman})  works
reasonably well with MT13+KW and AV18+KW interactions and for the
 \=pd  scattering states that have spin-uncoupled asymptotic channels: $J^\pi$=$1/2^-$,
$J^\pi$=$3/2^-$ and $J^\pi$=$5/2^+$ on Table  \ref{Tab_Trueman}.
Likewise for \=pp case, the  S-waves  require second order corrections, whereas
for P-wave the first order results are already accurate to four significant digits.
However  we have found that for the spin-coupled \=pd channels (like e.g.  P$_{1/2}$ and P$_{3/2}$), the standard, Coulomb-corrected, effective range formulae fails. 
Indeed, if the target nucleus is non-spherically symmetric, the asymptotic  \=p-A interaction  
acquires, beyond the Coulomb interaction, higher multipole terms, starting with the $1/r^3$ quadrupole ones. These terms are diagonal in total spin basis,
but due the fact that strong interaction does not conserve the total spin, the \=p-A states are a mixture of  different spin
states. Therefore  coupled \=pA channels remain asymptotically coupled by quadrupole interaction terms.
In this case, a modified effective range formulae must be derived and  expansion
(\ref{Trueman}) reformulated accordingly. This tedious task is  beyond the scope of this paper.

The validity of the Trueman relations, as an indirect way to obtain the antiproton-nucleus level shifts, is a crucial asset in view of  the
theoretical aspects of the PUMA project. 
Indeed, the computation of  \={p}-nucleus scattering lengths turns to be  a much easier task than determining the tiny energy shifts of the atomic levels.
If one could rely on this indirect approach to obtain the \={p}A level shifts, this will offer the  possibility to extended the  rigorous  theoretical calculations
to more complex structures than deuteron. Using the deuteron nuclear target we have demonstrated that this is indeed the case. Nevertheless one should bare in mind that
these relations should be redefined if strong magnetic interaction terms are present, as it will happen for non-spherical nuclear targets.

\begin{table}[h!]
\begin{center}
\begin{footnotesize}
\begin{tabular}{r  r r r r}\hline
\multicolumn{5}{c}{MT13 +KW }\\
    & $a_0$  (fm)         &   $\Delta E_1$ (keV)     &$\Delta E_2$ (keV) & $\Delta E$ (keV) \\\hline
$^2S_{1/2}$ n=1         &1.596-0.8569i   &  2.463-1.322i  & 2.259-1.014i  & 2.251-1.004i \\
 $^4S_{3/2}$ n=1        &1.647-0.8419i   &  2.541-1.299i  & 2.316-0.987i  & 2.321-0.984i \\
                        &   $a_1$ (fm$^3$) &      $\Delta E_1$ (meV)     & $\Delta E_2$ (meV) & $\Delta E$ (meV)      \\\hline
  $^4P_{5/2}$ n=2       &  0.450-2.68i   & 34.8-207i  & 34.8-207i &   26.2-215i \\
\multicolumn{5}{c}{AV18 +KW }\\
     & $a_0$  (fm)         &   $\Delta E_1$ (keV)     &$\Delta E_2$ (keV) & $\Delta E$ (keV) \\\hline
$^2S_{1/2}$ n=1        &1.505-0.8779i   &  2.323-1.355i  & 2.155-1.057i  & 2.147-1.044i \\
 $^4S_{3/2}$ n=1       &1.59-0.8771i    &  2.541-1.354i  & 2.257-1.039i  & 2.218-1.075i \\
                       & $a_1$ (fm$^3$) &     $\Delta E_1$ (meV)    &$\Delta E_2$ (meV) & $\Delta E$ (meV)      \\\hline
   $^4P_{5/2}$ n=2     &  0.469-2.57i        & 36.4-199i  &   36.4-199i &   39.9-204i\\
\end{tabular}
\caption{Atomic level shifts, calculated from \=pd scattering
lengths ($a_0$ and $a_1$) employing Trueman relations at first
order ($\Delta E_1$) and second order ($\Delta E_2$)   are
compared with the values obtained from direct binding energy
calculations ($\Delta E$).}\label{Tab_Trueman}
\end{footnotesize}
\end{center}
\end{table}

\section{Concluding remarks}\label{Conclusion}

We have presented In this letter  the first rigorous calculation of
the antiproton-deuteron states based on a coupled
(\={p}pn)-(\={n}nn) three-body system interacting with realistic NN and
\={N}N optical models.

The complex energies of the low lying states as well as the
corresponding scattering lengths and volumes have been obtained
for several combinations of NN and \={N}N models. The S-wave
energy shifts of atomic \=pd energy levels are model independent
but are not compatible with the measured quantities by a factor
two. The corresponding widths fall inside the (large) experimental
error bars. On the contrary, the P-waves are strongly model dependent
and does not tolerate  any approximations in the solution of the 3-body problem.
Such a dependence may be due  to the presence of a nearthreshold
singularities in the P-waves amplitudes,  making the system very
sensitive to small changes in the theoretical input:  they can easily transform a loosely bound P-wave 3-body bound state into a resonance
with the corresponding change of sign in the \={p}d level shitfs. 

The \={p} annihilation densities for an S- and P-sate have been
computed and found to follow closely the deuteron density up to
5-10 fm. A substantial part of the annihilation process, compared
to the \={p}p case, takes part in the tail part of deuteron.

We have investigated the validity of the Trueman relation,
established for point like particle scattering, to the case of a
composite target and found it to be accurate enough at the second
order for spin-uncoupled channels. For the spin-coupled channels
the Trueman relation should be generalized to account for the possible 
coupling in the asymptotic channel induced by $1/r^3$ quadrupole terms.

The actual goal of the PUMA experiment is to study the nuclear
surface densities using antiproton-nucleus annihilation. It is
assumed that  a substantial part of the annihilation process takes
place in the outer shell of the nucleus and that hey can be used
to probe the peripheral matter density. This hypothesis is
essentially confirmed by our results on deuteron. The presented
\={p}d annihilation densities  shows that, although for S-wave the
antiproton is able to penetrate deeply inside the deuteron, an
important fraction of annihilations takes place at large distance
from the center. This is even more visible for P-waves where the
annihilation process is displaced toward the nuclear periphery. In
any case, the annihilation density follows closely the asymptotic
density profile of the deuteron. This fact provides strong support
for the  major hypothesis of the PUMA experiment, as most of the
relevant annihilation signal is expected to happen from high angular momentum orbitals.

\section*{Acknowledgements}
This work was supported by french CNRS/IN2P3 for a theory project
"Neutron-rich light unstable nuclei". We would like to thank Prof. S. Wycech for helpful remarks and
express our gratitude to Prof. Johann Haidenbauer for his assistance in implementing the chiral EFT  \={N}N interaction model
\cite{Haidenbauer_JHEP_2017} and for readjusting the interaction parameters to match non-relativistic kinetic energy operators.

\end{document}